\documentclass{aastex6}

\begin{document}

\title{High-contrast Polarimetry Observation of T Tau Circumstellar Environment}

\author{
Yi YANG\altaffilmark{1,2},
Satoshi MAYAMA\altaffilmark{34},
Saeko S. HAYASHI\altaffilmark{1,2},
Jun HASHIMOTO\altaffilmark{3},
Roman RAFIKOV\altaffilmark{32,25},
Eiji AKIYAMA\altaffilmark{2},
Thayne CURRIE\altaffilmark{4},
Markus JANSON\altaffilmark{27},
Munetake MOMOSE\altaffilmark{29},
Takao NAKAGAWA\altaffilmark{10},
Daehyeon OH\altaffilmark{31,1,2},
Tomoyuki KUDO\altaffilmark{4},
Nobuhiko KUSAKABE\altaffilmark{3},
Lyu ABE\altaffilmark{6},
Wolfgang BRANDNER\altaffilmark{7},
Timothy D. BRANDT\altaffilmark{33},
Joseph C. CARSON\altaffilmark{9},
Sebastian EGNER\altaffilmark{28},
Markus FELDT\altaffilmark{7},
Miwa GOTO\altaffilmark{11},
Carol A. GRADY\altaffilmark{12,13,14},
Olivier GUYON\altaffilmark{4,3,30},
Yutaka HAYANO\altaffilmark{1,2,4},
Masahiko HAYASHI\altaffilmark{1,2}
Thomas HENNING\altaffilmark{7},
Klaus W. HODAPP\altaffilmark{15},
Miki ISHII\altaffilmark{2},
Masanori IYE\altaffilmark{2},
Ryo KANDORI\altaffilmark{2},
Gillian R. KNAPP\altaffilmark{8},
Jungmi KWON\altaffilmark{10},
Masayuki KUZUHARA \altaffilmark{3},
Taro MATSUO\altaffilmark{16},
Michael W. MCELWAIN\altaffilmark{12},
Shoken MIYAMA\altaffilmark{17},
Jun-Ichi MORINO\altaffilmark{2},
Amaya MORO-MARTIN\altaffilmark{8,18},
Tetsuo NISHIMURA\altaffilmark{4},
Tae-Soo PYO\altaffilmark{1,4},
Eugene SERABYN\altaffilmark{19},
Takuya SUENAGA\altaffilmark{1,2},
Hiroshi SUTO\altaffilmark{2,3},
Ryuji SUZUKI\altaffilmark{2},
Yasuhiro H. TAKAHASHI\altaffilmark{2,5},
Michihiro TAKAMI\altaffilmark{20},
Naruhisa TAKATO\altaffilmark{1,4},
Hiroshi TERADA\altaffilmark{2},
Christian THALMANN\altaffilmark{21},
Edwin L. TURNER\altaffilmark{8,22},
Makoto WATANABE\altaffilmark{23},
John WISNIEWSKI\altaffilmark{26},
Toru YAMADA\altaffilmark{24},
Hideki TAKAMI\altaffilmark{1,2},
Tomonori USUDA\altaffilmark{1,2},
Motohide TAMURA\altaffilmark{5,3,2}
}

\affil{
1 Department of Astronomical Science, SOKENDAI (The Graduate University for Advanced Studies), 2-21-1 Osawa, Mitaka, Tokyo 181-8588, Japan\\
2 National Astronomical Observatory of Japan (NAOJ), National Institutes of Natural Sciences (NINS), 2-21-1 Osawa, Mitaka, Tokyo 181-8588, Japan\\
3 Astrobiology Center, NINS, 2-21-1, Osawa, Mitaka, Tokyo 181-8588, Japan\\
4 Subaru Telescope, NAOJ, NINS, 650 North A'ohoku Place, Hilo, HI 96720, USA\\
5 Department of Astronomy, The University of Tokyo, 7-3-1 Hongo, Bunkyo-ku, Tokyo 113-0033, Japan\\
6 Laboratoire Lagrange (UMR 7293), Universite de Nice-Sophia Antipolis, CNRS, Observatoire de la Coted'azur, 28 avenue Valrose, 06108 Nice Cedex 2, France\\
7 Max Planck Institute for Astronomy, K{\"o}nigstuhl 17, 69117 Heidelberg, Germany\\
8 Department of Astrophysical Science, Princeton University, Peyton Hall, Ivy Lane, Princeton, NJ 08544, USA\\
9 Department of Physics and Astronomy, College of Charleston, 58 Coming St., Charleston, SC 29424, USA\\
10 Institute of Space and Astronautical Science, Japan Aerospace Exploration Agency, 3-1-1 Yoshinodai, Chuo-ku, Sagamihara, Kanagawa 252-5210, Japan\\
11 Universit{\"a}ts-Sternwarte M{\"u}nchen, Ludwig-Maximilians-Universit{\"a}t, Scheinerstr. 1, 81679 M{\"u}nchen, Germany\\
12 Exoplanets and Stellar Astrophysics Laboratory, Code 667, Goddard Space Flight Center, Greenbelt, MD\\
20771, USA\\
13 Eureka Scientific, 2452 Delmer, Suite 100, Oakland CA 96002, USA\\
14 Goddard Center for Astrobiology, NASA Goddard Space Flight Center, Greenbelt, MD 20771, USA\\
15 Institute for Astronomy, University of Hawaii, 640 N. A'ohoku Place, Hilo, HI 96720, USA\\
16 Department of Earth and Space Science, Graduate School of Science, Osaka University, 1-1 Machikaneyamacho, Toyonaka, Osaka 560-0043, Japan\\
17 Hiroshima University, 1-3-2 Kagamiyama, Higashihiroshima, Hiroshima 739-8511, Japan\\
18 Department of Astrophysics, CAB-CSIC/INTA, 28850 Torrej\'on de Ardoz, Madrid, Spain\\
19 Jet Propulsion Laboratory, California Institute of Technology, M/S 171-113 4800 Oak Grove Drive Pasadena, CA 91109 USA\\
20 Institute of Astronomy and Astrophysics, Academia Sinica, P.O. Box 23-141, Taipei 10617, Taiwan\\
21 Swiss Federal Institute of Technology (ETH Zurich), Institute for Astronomy,\\
Wolfgang-Pauli-Strasse 27, CH-8093 Zurich, Switzerland\\
22 Kavli Institute for Physics and Mathematics of the Universe, The University of Tokyo, 5-1-5 Kashiwanoha, Kashiwa, Chiba 277-8568, Japan\\
23 Department of Cosmosciences, Hokkaido University, Kita-ku, Sapporo, Hokkaido 060-0810, Japan\\
24 Astronomical Institute, Tohoku University, Aoba-ku, Sendai, Miyagi 980-8578, Japan\\
25 Astrophysics Department, Institute for Advanced Study, Princeton, NJ 08540, USA\\
26 H. L. Dodge Department of Physics \& Astronomy, University of Oklahoma, 440 W Brooks St., Norman, OK 73019, USA\\
27 Department of Astronomy, Stockholm University, AlbaNova University Center, SE-106 91 Stockholm, Sweden\\
28 European Southern Observatory, Karl Schwarzschildstr. 2, Garching 85748, Germany\\
29 College of Science and Center for Astronomy, Ibaraki University, 2-1-1 Bunkyo, Mito 310-8512, Japan\\
30 Steward Observatory, University of Arizona, 933 N Cherry Ave., Tucson AZ 85719, USA\\
31 National Meteorological Satellite Center, 64-18 Guam-gil, Gwanghyewon-myeon, Jincheon-gun, Chungbuk, South Korea\\
32 Department of Applied Mathematics and Theoretical Physics, Centre for Mathematical Sciences, University of Cambridge, Wilberforce Road, Cambridge CB3 0WA, UK\\
33 Department of Physics, Broida Hall, University of California, Santa Barbara, CA 93106-9530\\
34 SOKENDAI(The Graduate University for Advanced Studies), Shonan International Village, Hayama-cho, Miura-gun, Kanagawa 240-0193, Japan
}

\email{yi.yang@nao.ac.jp}

\begin{abstract}
We conducted high-contrast polarimetry observations of T Tau in the H-band, using the HiCIAO instrument mounted on the Subaru Telescope, revealing structures as near as 0.$\arcsec$1 from the stars T Tau N and T Tau S. The whole T Tau system is found to be surrounded by nebula-like envelopes, and several outflow-related structures are detected in these envelopes. We analyzed the detailed polarization patterns of the circumstellar structures near each component of this triple young star system and determined constraints on the circumstellar disks and outflow structures. We suggest that the nearly face-on circumstellar disk of T Tau N is no larger than 0.$\arcsec$8, or 117 AU, in the northwest, based on the existence of a hole in this direction, and no larger than 0.$\arcsec$27, or 40 AU, in the south. A new structure ``N5" extends to about 0.$\arcsec$42, or 59 AU, on the southwest of the star, believed to be part of the
disk. We suggest that T Tau S is surrounded by a highly inclined circumbinary disk with a radius of about 0.$\arcsec$3, or 44 AU, with a position angle of about 30$^\circ$, that is misaligned with the orbit of the T Tau S binary. After analyzing the positions and polarization vector patterns of the outflow-related structures, we suggest that T Tau S should trigger the well-known E-W outflow, and is also likely to be responsible for a southwest precessing outflow ``coil" and a possible south outflow.

\end{abstract}


\section{Introduction}

As a prototype of the large class of T Tauri pre-main sequence stars, T Tau has attracted considerable attention from astronomers studying star and planet formation processes. T Tau is located about 146.7$\pm$0.6 pc from us \citep{Loinard2007} and has an age of about 1--2 Myr \citep{Kenyon1995}. It is actually a triple system, consisting of the north single star T Tau N and the south binary T Tau Sa/Sb. T Tau N, which has a mass of about 1.95 $M_\odot$ \citep{Kohler2016}, is believed to be a class II young stellar object (YSO) \citep{Furlan2006,Luhman2010}. T Tau S, located about 0.$\arcsec$7 south of T Tau N, was first discovered by \citet{Dyck1982} and could be a class I YSO \citep{Furlan2006,Luhman2010}. \citet{Koresko2000} discovered that T Tau S is actually a binary Sa/Sb with a separation of about 0.$\arcsec$1, with masses of 2.12$\pm$0.10 $M_\odot$ for Sa and 0.53$\pm$0.06 $M_\odot$ for Sb \citep{Kohler2016}. The orbit of T Tau Sa/Sb has a semi-major axis of 12.5$^{+0.6}_{-0.3}$ AU and an eccentricity of 0.56$^{+0.07}_{-0.09}$. The orbit of the T Tau N--S system is not well constrained, but is likely to have a semi-major axis of 430$^{+790}_{-250}$ AU and an eccentricity of 0.7$^{+0.2}_{-0.4}$ \citep{Kohler2016}.

Although T Tau has been studied extensively, there are still some important issues that need to be resolved for this system. The first issue is the circumstellar disk structures in the system. \citet{Akeson1998} made 3-mm continuum observations of T Tau N and estimated that it might have a nearly face-on disk with an outer radius of about 41 AU, but their beam-size ($0.\arcsec59\times0.\arcsec39$) was not sufficient to reveal the details. Other research has given quite different results, such as \citet{Gustafsson2008}, which estimated the outer radius 85--100 AU, corresponding to about 0.$\arcsec$6--0.$\arcsec$7, based on a spectral energy distribution (SED) simulation, while \citet{Podio2014} derived a 110-AU size based on CN 5-4 line observations. T Tau S binary has an extinction of $A_V=15$ \citep{Duchene2005}, much higher than that of T Tau N ($A_V=1.95$, \citet{Kenyon1995}). This difference has been attributed to a compact edge-on circumbinary disk around T Tau S \citep{Duchene2005} or blockage of the light due to the circumstellar disk around T Tau N \citep{Hogerheijde1997, Beck2001}. Besides the possible edge-on circumbinary disk, T Tau Sa may be surrounded by an edge-on circumstellar disk of radius 2--3 AU \citep{Duchene2005} oriented north--south, while the circumstellar disk around T Tau Sb may not be far from face-on, based on middle infrared interferometry observations and SED simulations \citep{Ratzka2009}. To date, discussions of the disk structures in this system have mainly been based on indirect measurements. A direct observation of the disks in this system with sufficiently high resolution is crucial for understanding the disk system.

The second issue is the sources of the outflows. Several outflows have been reported in previous observations of this system. \citet{Bohm1994} observed T Tau with a spectrograph mounted on the 2.2-m telescope at the Calar Alto Observatory and reported the discovery of an east--west outflow (hereafter, the E-W outflow) and a southeast--northwest outflow (hereafter, the northwest outflow) from this system. However, it is still under debate which stars trigger which outflow. Some research, e.g., \citet{Bohm1994} and \citet{Gustafsson2010}, attributed to T Tau N the E-W outflow and to T Tau S the northwest outflow, but \citet{Ratzka2009} claimed that T Tau S is responsible for the E-W outflow. In addition, \citet{Gustafsson2010} suggested the presence of another southwest outflow coming from T Tau Sb based on near infrared hydrogen emission observations. \citet{Kasper2016} also detected a coil-like structure extending to the southwest of the T Tau system in their multi-band near infrared high-contrast imaging study. This structure can be attributed to a precessing outflow, but it is still unknown whether this originates from T Tau N or S. Identification of these outflow features will help determine the evolution of the stars in the multiple system.

Near infrared high-contrast polarimetry imaging is sensitive to the disk and envelope structures around the young binaries, and can reach a spatial resolution smaller than 0.$\arcsec$1. It has been used to resolve several protoplanetary disks (e.g., \citet{Oh2016}). \citet{Whitney1993} and \citet{Murakawa2010} have simulated the polarized image of young stellar objects with bipolar outflow cavities, showing that for edge-on disks with outflow cavities, the vectors near the disk plane tend to be aligned with the disk. Therefore, through polarimetry imaging it is possible to detect the outflow cavity structures from the directions of the vectors, and distinguish the disk and envelope.


In this paper, we will present the results of near infrared high contrast polarimetry observations of the T Tau system, and from the results we constrain the disk size around T Tau N and S. Section 2 presents the observations and data reduction methods. In Section 3, the observation results are introduced. Section 4 discusses circumstellar disks and outflows based on the observation results. In Section 5, conclusions are made.

\section{Observations and Data Reduction}

Observations of T Tau, part of the Strategic Explorations of Exoplanets and Disks with Subaru (SEEDS) survey started in 2009, was conducted on January 8th 2015, Hawaii Standard Time, using the High Contrast Instrument for the Subaru Next Generation Adaptive Optics (HiCIAO, \citet{Tamura2006}) and the adaptive optics (AO) system AO188 \citep{Hayano2010} mounted on the 8.2-m Subaru Telescope. The observations were made in the H-band, and the quad-polarized differential imaging (qPDI) mode was used in this observation. In this mode, a double-Wollaston prism is used to split the incident light into four 512$\times$512-pixel channels, corresponding to two o-polarization and two e-polarization channels. The detector has a pixel scale of 9.50 mas pixel$^{-1}$, and the AO system helps to limit the full width at half maximum (FWHM) of the stellar point-spread function (PSF) to 0.$\arcsec$07. To calibrate the Stokes parameters, a half-wave plate was rotated among four angles: $0^{\circ}$, $45^{\circ}$, $22.5^{\circ}$, and $67.5^{\circ}$. Finally, 36 frames, each with an exposure time of 5 seconds and four coadds, were collected, corresponding to a total integration time of 12 minutes. 

The data were reduced using the Image Reduction and Analysis Facility (IRAF) pipeline. Generally, the reduction steps include correction of stripes and the flat field, removal of bad pixels and distortions, and generation of a polarized intensity (PI) image. To obtain the PI image, the Stokes $+Q$, $+U$, $-Q$, and $-U$ images were first obtained by subtracting the e-images from the o-images, and then the formulas $Q=((+Q)-(-Q)/2$ and $U=((+U)-(-U))/2$ were used to construct the Stokes $Q$ and Stokes $U$ images. After correcting the instrumental polarization, the PI image could be obtained from the Stokes $Q$ and $U$ images using $PI=\sqrt{Q^2+U^2}$. Through this reduction process, the unpolarized light from the stars is removed, leaving only the polarized light from the circumstellar disk or envelopes. The Stokes $I$ image, or intensity image, which contains both polarized and unpolarized light, was obtained by averaging the sum of the o- and e-images in all frames.

\section{Results}

The PI image of T Tau is presented in Figure 1, showing an area of about $4.\arcsec9\times4.\arcsec9$ around this triple star system. This is one of the highest resolution polarization images of T Tau ever made and has revealed a number of new structures not previously reported. All three stars in this system, T Tau N and T Tau Sa/Sb, as well as the nebula-like structures surrounding them, can be seen in this PI image. We measured the distance between T Tau Sa and Sb to be about $0.\arcsec10\pm0.\arcsec01$, and their position angle to be about $342^{\circ}\pm1^{\circ}$ (measured from north to east). These results are consistent with those obtained for December 2014 and January 2015 by \citet{Kohler2016}, which indicated that the T Tau Sa/Sb system has a separation of 0.$\arcsec$11 and a position angle of about $345^{\circ}$ in the Br$\gamma$ and Ks bands. T Tau N is saturated within about 0.$\arcsec$1 so we use a soft mask to block the saturated part. Even though an accurate determination of the stellar position is not straightforward, the circular shape of the saturation allows us to reasonably estimate its central location within 0.$\arcsec$01. Using the center of the saturated circle as the position of T Tau N with a 0.$\arcsec$01 uncertainty, the distance between T Tau N and Sa is estimated to be $0.\arcsec68\pm 0.\arcsec01$, consistent with the result of \citet{Kohler2016} for December 2014 (0.$\arcsec$69) in the Br$\gamma$ band. \citet{Kohler2016} determined that the orbit of T Tau N--S has an inclination of about 52$^\circ$ and a position angle of the ascending node of about 156$^\circ$. Since the position angle of T Tau S relative to T Tau N is about 195$^\circ$ from the PI image, T Tau S is farther from us than T Tau N.

Since the light scattered by the dust around the stars is polarized while the light direct from the stars is non-polarized, polarimetry observations can readily resolve the circumstellar disks and envelopes near the stars. Several petal-like nebular structures that appear to extend from the T Tau stars are observed, nearly filling the whole field of view.

The polarization vector maps of the system are shown in Figure 2. The vectors show the polarization position angles $\theta_p$, which are calculated using the formula $\theta_p=0.5{\rm tan}^{-1}(U/Q)$. It can be seen when the vectors are plotted over these structures that they are centrosymmetric around either T Tau Sa/Sb or T Tau N, implying that they are real structures associated with these point sources (e.g., \citet{Whitney1992}). After comparing them with the results of previous observations, these surrounding structures are believed to be the inner part of the envelopes surrounding the triple system, as suggested by \citet{Mayama2006}, as well as the circumstellar disk around T Tau N and the circumbinary disk around T Tau S. A detailed discussion will be presented in section 4.1. 

The PI image contains certain dark regions, such as a clear large hole in the northwest with an opening angle of about 40$^\circ$ from T Tau N. These areas do not exhibit reflected polarized light, but this does not necessarily mean that there is no scattered light or scattering structures (e.g., \citet{Perrin2009}).  According to \citet{Perrin2009}, a fake gap is expected to be caused by backscattered light on the far side of an inclined disk. This is not the case for T Tau, however, as T Tau N has a nearly face-on disk and its surrounding envelopes are not likely to cause backscattered light. Therefore, we suggest that this hole reflects a real structure.  In addition, a J-band T Tau image obtained by the Canada France Hawaii Telescope (CFHT)\citep{Roddier1999}\footnote{A false color image is viewable at http://www.cfht.hawaii.edu/Science/Astros/Imageofweek/ciw290500.html} also shows a dark region located in the northwest of the nebular structure and its position is near this hole. However, these data are very old so further observations are needed.

In the eastern area, there are two clear dark lines at position angles of 50$^\circ$ and 100$^\circ$, which have not been reported in any previous observations. There are also dark regions in the southern and western areas; the south area is about 1.$\arcsec$8 from T Tau N, while the western one is about 2.$\arcsec$2 from T Tau N. In the H-band image by \citet{Mayama2006}, the southern and western boundaries of the nebular structures detected in the H-band are about 2$\arcsec$ from T Tau N, which is generally consistent with our results. This suggests that these dark regions should represent the boundaries of the envelope.

In the nebular structures, especially in the southwestern area, there are complicated structures, shown in red in the figures. For clarification, they are labeled N1--N5 and ``coil" in Figure 1(b). To help understand what these structures are, we made detailed comparisons with previous observations, including the multi-band (J, H, K, and several hydrogen emission lines) observations in 2014 and 2015 using VLT/SPHERE-IRDIS, IRDIFS \citep{Kasper2016}, the $H_2\ v=1-0\ S(1)$ emission line observations performed in 2004 using VLT/SINFONI \citep{Gustafsson2010}, and the $H_2\ v=1-0\ S(1)$ emission line observations using VLT/NACO with a Fabry--Perot interferometer \citep{Herbst2007}. The comparisons were done by carefully overlapping these images on our image, matching the positions of the stars (for T Tau S, we use the barycenter as a reference) and scales, to see the differences of the structures. A brief summary of the comparison results is given in Table 1. 

In our PI image, the N1 structure is associated with T Tau S, and extends to about 0.$\arcsec$5 southeast from T Tau Sa. Its position is near the R1 structure reported by \citet{Kasper2016}, the 1 and 2 structure reported by \citet{Gustafsson2010}, and the C4 structure reported by \citet{Herbst2007}, suggesting that they are related. 

The position of the N2 structure appears to match the position of the R2 structure reported by \citet{Kasper2016}, but in our H-band image it looks much shorter (about 0.$\arcsec$1) than the R2 in the J band image of \citet{Kasper2016} (about 0.$\arcsec$25). The N3 structure is about 0.$\arcsec$4 from T Tau Sa and has a length of about 0.$\arcsec$4. Its position also overlaps well with the position of the R3 structure in the J-band data of \citet{Kasper2016}. It is about 0.$\arcsec$2 below structure 3 reported by \citet{Gustafsson2010} and the C3 structure reported by \citet{Herbst2007}.  

The N4 structure in our image appears more extended than the R4 structure in \citet{Kasper2016}; its brightest part has a length of about 0.$\arcsec$5 and width of about 0.$\arcsec$2. This feature covers the positions of the R4 structure reported by \citet{Kasper2016}, structures 5 and 6 reported by \citet{Gustafsson2010}, the C1 and C2 structures reported by \citet{Herbst2007}, and a knot-like structure ``N1" reported by \citet{Saucedo2003} in their HST/STIS Ly$\alpha$ observation. Also, it seems to be connected with N3, N5, and the coil in our image. 

N5 appears to start at T Tau N and extends to about 0.$\arcsec$4, or 59 AU, and PA $\sim$220$^\circ$ from T Tau N. This structure does not have any counterparts reported in previous observations and hence it is a newly detected structure. Although it looks like it is connected with N4, the polarization vectors plotted over N5 are centrosymmetric with respect to T Tau N, while the vectors for N4 are centrosymmetric with respect to T Tau S. This implies that N5 and N4 are illuminated by different sources and are therefore different structures. \citet{Ray1997} suggested a structure T Tau R near T Tau N, and \citet{Csepany2015} reported a tentative object 144 milliarcseconds south of T Tau N. The position of N5 does not fit the object reported by \citet{Csepany2015}. If T Tau R is an object orbiting T Tau N, its position angle should have changed from about 44$^\circ$ to about 240$^\circ$ over the past 22 years, considering its location to be 0$\arcsec$.3, or 44 AU, from T Tau N. As a result, N5 is also unlikely to be related to T Tau R. In addition, the vectors plotted over N5 surround only T Tau N, and thus another luminous object near it can be excluded. Therefore, we conclude that N5 could be part of the circumstellar disk around T Tau N. 

A coil structure, first reported by \citet{Kasper2016}, is clearly seen in our PI image. Its appearance is similar to the J-band structure observed by \citet{Kasper2016}. In our image, it looks like it can be connected with the N4 structure, and extends to about 2.$\arcsec$1, or 294 AU, with a position angle of about 230$^\circ$ from T Tau N. The T Tau NW structure, which was reported by \citet{Herbst2007} and could be related to the northwest outflow, is not found in our image. After comparing with the results of \citet{Kasper2016}, it is likely that this structure is just at the northwestern edge of our field of view and cannot be seen in our image. 

\floattable
\begin{deluxetable}{cccc}\tablecaption{Comparison of outflow structures}
\tablecolumns{4}
\tablenum{1}
\tablewidth{0pt}
\tablehead{
\colhead{This paper} &
\colhead{\citet{Herbst2007}} &
\colhead{\citet{Gustafsson2010}} &
\colhead{\citet{Kasper2016}} 
}
\startdata
N1 & C4 & 1a/b, 2 & R1  \\
N2 & \nodata & \nodata & R2  \\
N3 & C3 & 3 & R3   \\
N4 & C1, C2 & 5, 6 &  R4 \\
\enddata
\tablecomments{This table lists the structures detected in our image and possible related structures in previous observations.}
\end{deluxetable}

\section{Discussion}

\subsection{Circumstellar Disks}

In this section, we focus on the circumstellar disks in the T Tau system. \citet{Akeson1998} reported that T Tau N might have a nearly face-on disk with an outer radius of about 41 AU, and previous SED fitting \citep{Gustafsson2008} and CN line observations (e.g., \citet{Podio2014}) have suggested that it has a circumstellar disk of about 100 AU. A disk of this scale should be detectable within the field view of our observation, although it is not easy to distinguish from the envelope. The structures around T Tau N close to 0.11$\arcsec$ (16 AU) may correspond to the previously reported circumstellar disks. 

The circumstellar disk of T Tau N should not exceed the inner boundary of the ``hole" to the northwest of T Tau N of about 0.$\arcsec$8, or 117 AU. However, in the south, the vector map in Figure 2 shows that only the vectors with distances smaller than $\sim$0.$\arcsec$27, or 40 AU, are centrosymmetric with respect to T Tau N. As described in Section 3, N4 and N5 should be considered independent structures, and structure N5 is likely to be related to T Tau N, because the vectors on it are centrosymmetric with respect to T Tau N. Even if N5 is part of the disk, its outer size will be only about 59 AU.

 It would be strange for the disk to be so asymmetric. In addition, \citet{Miranda2015} showed that for an equal-mass binary, the circumstellar disk size decreases as the orbital eccentricity increases. The mass ratio $q$ ($q=M_b/(M_a+M_b)$, where $M_a$ and $M_b$ represent the masses of the binary primary star and secondary star, respectively) of the T Tau N--S system is about 0.43, not far from 0.5, and based on the best orbital parameters derived by \citet{Kohler2016}, namely that the N--S orbit has a semi-major axis of about 430 AU and an eccentricity about 0.7, the circumstellar disk size should be at least smaller than 0.2 times the orbital semi-major axis, i.e., 86 AU. In this case, a disk with a radius of about 117 AU is obviously too large. It is possible that either the boundary of the envelope is at 117 AU, i.e., the actual radius of the disk is much smaller than 117 AU, or the N--S orbit parameters derived by \citet{Kohler2016} are not accurate (e.g., the semi-major axis of the N--S orbit derived by \citet{Kohler2016} ranges from 180 AU to 1220 AU) and the real orbit may have a larger semi-major axis and smaller eccentricity. At any rate, based on our observation results, the disk in the south of T Tau N extends to no more than 0.$\arcsec$27, or 40 AU, while in the northwest, its size should be smaller than 0.$\arcsec$8, or 117 AU.

For T Tau S, some vectors, such as those to the northeast and southwest of the stars, appear to be aligned in one direction with position angles of about 30/210$^\circ$, as indicated by the green arrow in Figure 2 (b), while the vectors to the southeast and northwest of the stars show centrosymmetric characteristics, they still have a tendency of alignment with a position angle of about 30/210$^\circ$. This ``aligned'' centrosymmetry is consistent with the presence of a ``polarization disk" (e.g., \citet{Tamura1991,Whitney1993,Murakawa2010}) and is considered to result from multiple scattering in dense regions of the disk. The light is scattered multiple times in an optically thick disk, but is scattered less in an optically thin region, like bipolar outflow cavities. According to the simulation results of \citet{Murakawa2010}, the vectors near the disk plane seem to be aligned with the disk's orientation, but the vectors in other areas, such as the outflow cavities, will still be centrosymmetric, as the light in those regions may only be scattered once. Considering that T Tau S itself is a class I object, this is a reasonable explanation. Also, an optically thick inclined disk will cause a large extinction, which is consistent with previous findings that T Tau S has a very large extinction. The rough range of the ``polarization disk" is consistent with the real disk size as suggested by \citet{Murakawa2010}, and based on this we can estimate that the disk should have a radius of about 0.3$\arcsec$, or 44 AU, with a position angle of about 30$^\circ$. The actual inclination of this circumbinary disk is hard to estimate from just the vector map, while in the simulation of \citet{Murakawa2010}, the aligned vectors near the disk plane appeared for H = 0.3 and $\theta_{inc}$ = 60$^\circ$ and with a small maximum dust size $\alpha_{max}$ = 0.25 $\mu$m. Considering that the actual maximum dust size in the disk around T Tau S could be larger, its inclination should also be considered to be larger, to align the vectors with the disk. Therefore, it seems that the inclination of the disk should be larger than 60$^\circ$. We hope future ALMA observations of the gas velocity around the disk will lead to a more accurate result.

Binaries will open central gaps in circumbinary disks, and for circumbinary disks which are coplanar with the binary orbital planes, the gap size is usually about 2--3 times the binary semi-major axis (e.g., \citet{Artymowicz1994}). For a circumbinary disk which is misaligned with the binary orbital plane, the opened gap size seems to be smaller (e.g., \citet{Miranda2015}). \citet{Miranda2015} calculated the gap size of circumbinary disks under different misaligned angles relative to the binary orbit and eccentricities (Figure 9 of \citet{Miranda2015}) around binaries with mass ratios of $q=$ 0.1, 0.3, and 0.5. Considering that the mass ratio of T Tau S is about 0.2, the opened gap size can be estimated from the $q=0.1$ and 0.3 cases. The orbit of T Tau S has a semi-major axis of about 13 AU, eccentricity of about 0.56, and inclination of about 20$^\circ$. If the circumbinary disk around T Tau S has an inclination larger than about 60$^\circ$ as we suggested, the misaligned angle between the disk and orbital plane is larger than about 40$^\circ$. Figure 9 of \citet{Miranda2015} shows that for a binary mass ratio of $q=0.1$ and 0.3, orbital eccentricity of 0.56 and disk misaligned angle of 45$^\circ$, the opened gap size is about 2.5 times of the binary semi-major axis, so it is reasonable to suggest that for $q=0.2$ and misaligned angle 40$^\circ$, the opened gap size should be close to 2.5 times the binary semi-major axis. \citet{Miranda2015} showed that the larger the inclination between the binary orbit and disk, the smaller the gap size that the binary opens. Therefore, the gap in the circumbinary disk of T Tau S should be smaller than 32.5 AU. 

\subsection{Outflows}
In this section, we briefly analyze the structures associated with the outflows. To help in this analysis, we briefly discuss the polarization vectors and the relative distance of T Tau N and S.

In general, the polarization vectors on real structures, e.g., circumstellar disks and envelopes, will be centrosymmetric around the central star. If we draw perpendicular vectors relative to the polarization vectors, then we will see that the vectors point to the central star. In binary/multiple star systems, this can help us judge which star illuminates the structures. This image is presented in Figure 3. From this image, we can see that for T Tau N the vectors with position angles from roughly $-$45$^\circ$ to 110$^\circ$ relative to T Tau N point to T Tau N, indicating that they are illuminated by T Tau N. For T Tau S, the vectors with position angles from roughly 135$^\circ$ to $-$45$^\circ$ point to T Tau S, indicating that they are illuminated by T Tau S. For the other areas of the envelopes, most of the vectors appear to point to the area between T Tau N and T Tau S, indicating that they are illuminated by both stars. Considering that T Tau S locates generally south (position angle $\sim195^\circ$) of T Tau N, we expect that the area illuminated by T Tau N and S are generally horizontal symmetric in Figure 3. However, in Figure 3, they are generally symmetric with position angle about $-45^\circ$. This may indicate some internal structures in the envelope, but further observations are necessary to confirm it.

As mentioned in Section 4.1, N5 appears to be part of the circumstellar disk around T Tau N. The other structures, N1--N4, as well as the coil, could be related to the outflows in this system. 
The T Tau S system, as discussed in Section 4.1, has a nearly edge-on disk oriented at about 30$^\circ$, and N1 and N4 are located at its opposite sides. Therefore, we suggest that N1 and N4 could represent two opposite outflow cavities of T Tau S.

Based on the simulation results reported by \citet{Murakawa2010}, the direction of the aligned vectors should be generally perpendicular to the direction of the bipolar outflow cavity. Therefore, from the vector map, we can estimate the position angles of the outflow cavities to be about 120$^\circ$ and 300$^\circ$. \citet{Eisloffel1998} derived the tangential velocity position angle for HH 155, an object located $\sim$30$\arcsec$ west of T Tau and believed to be caused by the E-W outflow, to be about 305$^\circ$, near the position angle we derived. Also, \citet{Kasper2016} noted a T Tau SE structure in their $H_2$ image, even though the signal-to-noise ratio was too low to confirm its presence. We notice that its position angle is 119$^\circ$, which is close to the position angle of the outflow cavities we derived. We suggest that this T Tau SE could be a real structure related to this near eastern--western outflow. In summary, we suggest that the well-known E-W outflow is triggered by T Tau S and N1 and N4 represent the bipolar outflow cavities related to this outflow.  

For the other structures, \citet{Gustafsson2010} suggested that their structure 3 could represent a southwest outflow, which we here refer to as a ``south" outflow, to distinguish from the southwest outflow corresponding to the coil. Since our N2 and N3 structures are located near their structure 3, their connection to the outflow is readily implied. The coil, as \citet{Kasper2016} suggested, is likely to be a precessing outflow, but its origin remains a mystery. There have been no previous reports on this structure \citet{Kasper2016}. A possible related report is that of a ``bow" structure by \citet{Gustafsson2010}. 

The average position angle of this coil is about 249$^\circ$ ($\sim237^\circ$--$261^\circ$) relative to T Tau S and 229$^\circ$ ($\sim221^\circ$-$237^\circ$) relative to T Tau N. The vectors in Figure 3 on the coil point towards T Tau S with an average position angle of about 250$^\circ$ (corresponding to an average polarization position angle of about 160$^\circ$) rather than T Tau N, indicating that the coil structure is illuminated by T Tau S. Since T Tau N is in front of T Tau S, if the coil is one of the outflow structures emitted by T Tau N, we expect to see the vectors in Figure 3 pointing to T Tau N rather than T Tau S. Thus we suggest that the coil is more likely to be related to the T Tau S system. In this case, there are three outflows that could be related to T Tau S, though only two stars have so far been detected in this system. A simple hypothesis is that there is another companion star that has not yet been detected. Another possibility is that some structures, such as N2 and N3, are not caused by an outflow but by an inflow. Future observations, especially high-resolution spectroscopy observations and high-resolution CO gas emission line observations like ALMA, will help resolve this problem.

In addition, the T Tau NW structure is possibly located at the northwest edge of the field of view near the big hole, so we suggest that this hole is likely to be caused by the northwest outflow. Since the vectors in Figure 3 of the envelope structures near the hole generally point to T Tau N, it is likely that the remaining northwest outflow is related to T Tau N. 

A summary of our interpretation regarding the outflows is listed in Table 2 and illustrated in Figure 4. There have been no previous reports on the relative distances of the outflow related structures N1--N5 and the coil. In our interpretations, N1--N4 and the coil should be near T Tau S, i.e., farther than T Tau N and N5. Knowing the velocities of these structures may help determine more accurate relative distances. In addition, just from near infrared polarimetry observations it is hard to judge which outflows are actually triggered by T Tau Sa or T Tau Sb. We hope ALMA observations can provide further information regarding this issue.

\floattable
\begin{deluxetable}{ccc}\tablecaption{Summary of outflows}
\tablecolumns{3}
\tablenum{2}
\tablewidth{0pt}
\tablehead{
\colhead{Star} &
\colhead{Structure} &
\colhead{Outflow} 
}
\startdata
T Tau S & N1, N4 & E-W  \\
T Tau S & N2, N3 & South  \\
T Tau S & Coil & Southwest  \\
T Tau N & e.g., \citet{Herbst2007} & Northwest \\
\enddata
\end{deluxetable}

\section{Conclusion}
Using the HiCIAO instrument mounted on the Subaru Telescope, we resolved the nearby structures around the T Tau triple star system and derived new insights into the disks and outflows in this system. From the $PI$ image we found that this triple system is surrounded by petal-like envelopes, and some structures in these envelopes could be associated with outflows from the stars. Considering the existence of a hole in the northwest, we suggest that the face-on disk of T Tau N should not exceed about 0.$\arcsec$8, or 117 AU, in this direction, and it is no more than about 0.$\arcsec$27, or 40 AU, in the south. Also, we discovered a new structure N5, extending to about 59 AU from T Tau N, which could be part of the disk around T Tau N. For T Tau S, the vector map implies that it has a highly inclined disk with a radius of about 44 AU with a position angle of about 30$^\circ$. 

We also investigated the sources of previously discovered outflows. We believe that T Tau S triggers the well-known E-W outflow, because it has a nearly north--south circumbinary disk, and the outflow related structures N1 and N4 are located opposite this outflow. Since the vectors on the coil in Figure 3 point to T Tau S rather than T Tau N, we conclude that the coil is located in the envelope around T Tau S and hence is more likely to be triggered by T Tau S than T Tau N. The N2 and N3 structures appear to represent an outflow extending to the south, though if this is the case there should be one more star, currently undetected, in the T Tau S system. Finally, we have detected a large hole to the northwest of T Tau N and the vectors in Figure 3 point to T Tau N, it is possible that the Northwest outflow is related to this hole.  

\acknowledgements{}
We thank the anonymous \textbf{reviewer} who has greatly helped us improve this paper. 
This study was based on data collected by the Subaru Telescope, which is
operated by the National Astronomical Observatory of Japan (NAOJ) and National Institutes of Natural Sciences (NINS). 
We thank the Subaru Telescope staff for their support during the observations. 
We would also like to acknowledge the access given to us of the SIMBAD database operated by the Strasbourg Astronomical Data Center (CDS), Strasbourg, France. 
IRAF is distributed by the National Optical Astronomy Observatory, which is operated by the Association of Universities for Research in Astronomy (AURA) under a cooperative agreement with the National Science Foundation. MT is supported by a Grant-in-Aid for Scientific Research (No.15H02063). JC is supported by the U.S. National Science Foundation under Award No. 1009203.
The authors wish to recognize and acknowledge the very significant cultural role and reverence that the summit of Maunakea has always had within the indigenous Hawaiian community. We are most fortunate to have the opportunity to conduct observations from this mountain. 

\bibliographystyle{apj}

\clearpage

\begin{figure}[ht!]
	\figurenum{1}
    \gridline{\fig{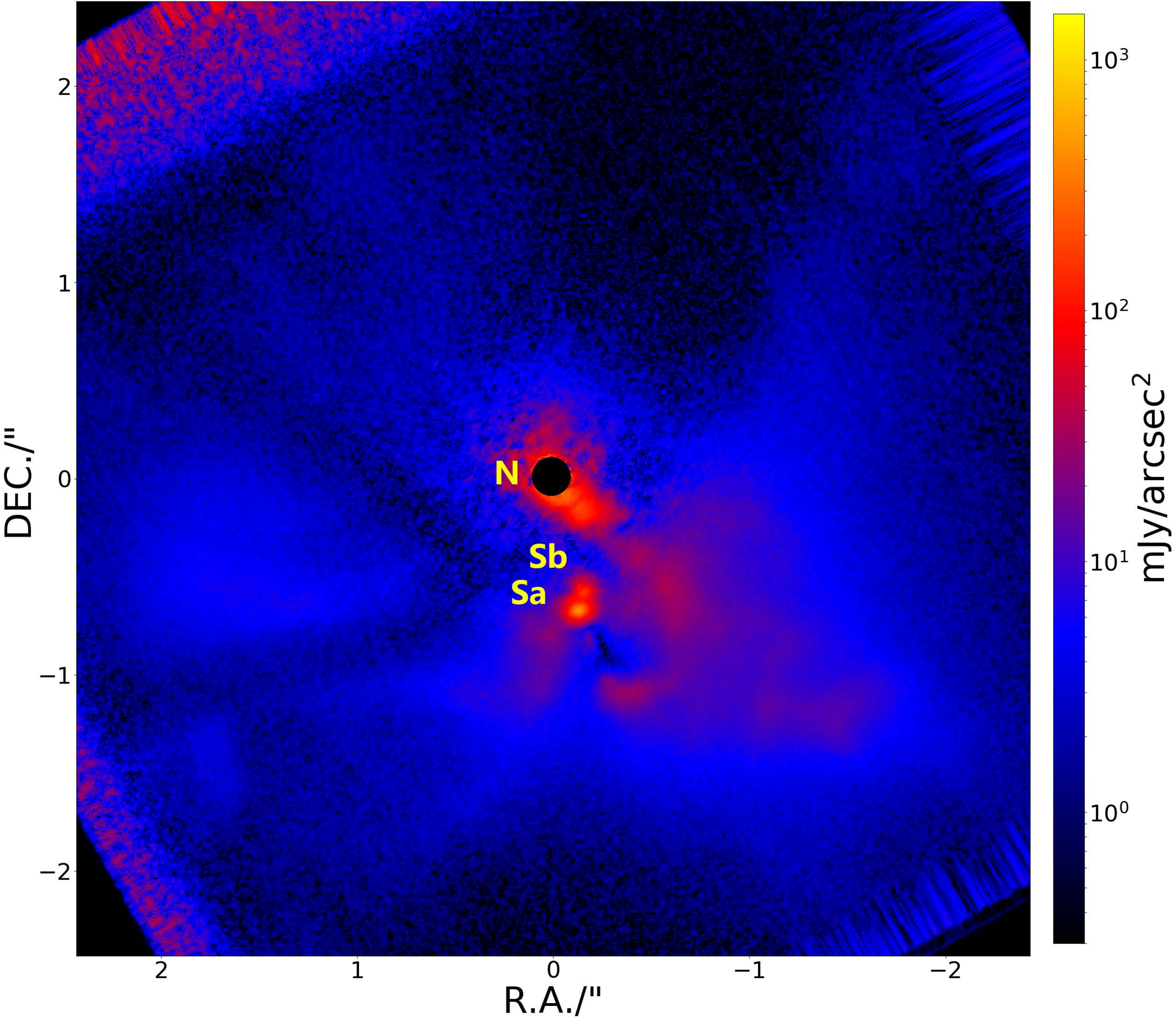}{0.5\textwidth}{a}
         \fig{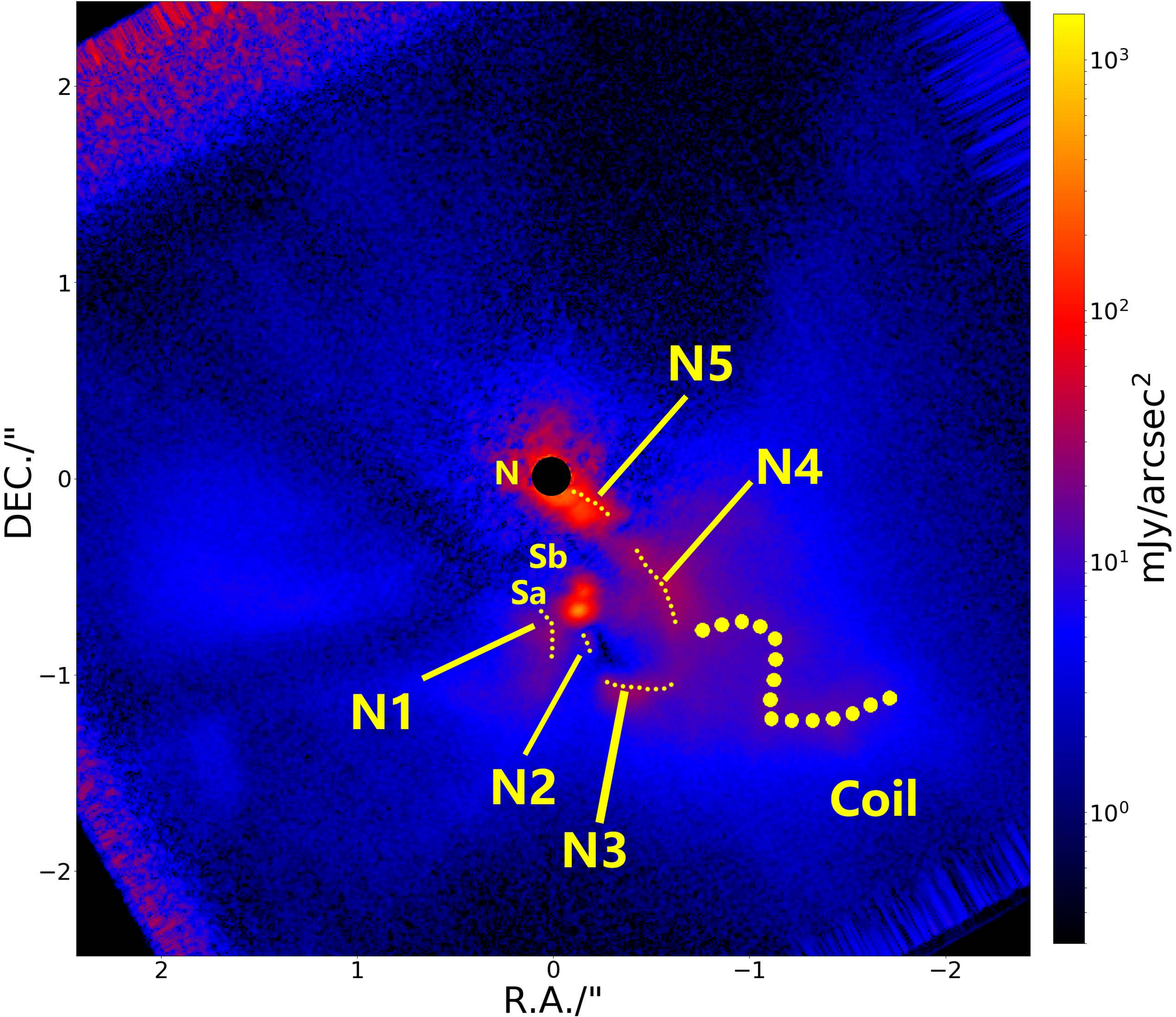}{0.5\textwidth}{b}
	}
	\caption{a: PI image of T Tau; b: as (a) but with the outflow-related structures labeled. A 0.$\arcsec$1-radius black circle is used in all the $PI$ images to block the saturated part. }
\end{figure}

\clearpage

\begin{figure}[ht!]
	\figurenum{2}
   \gridline{\fig{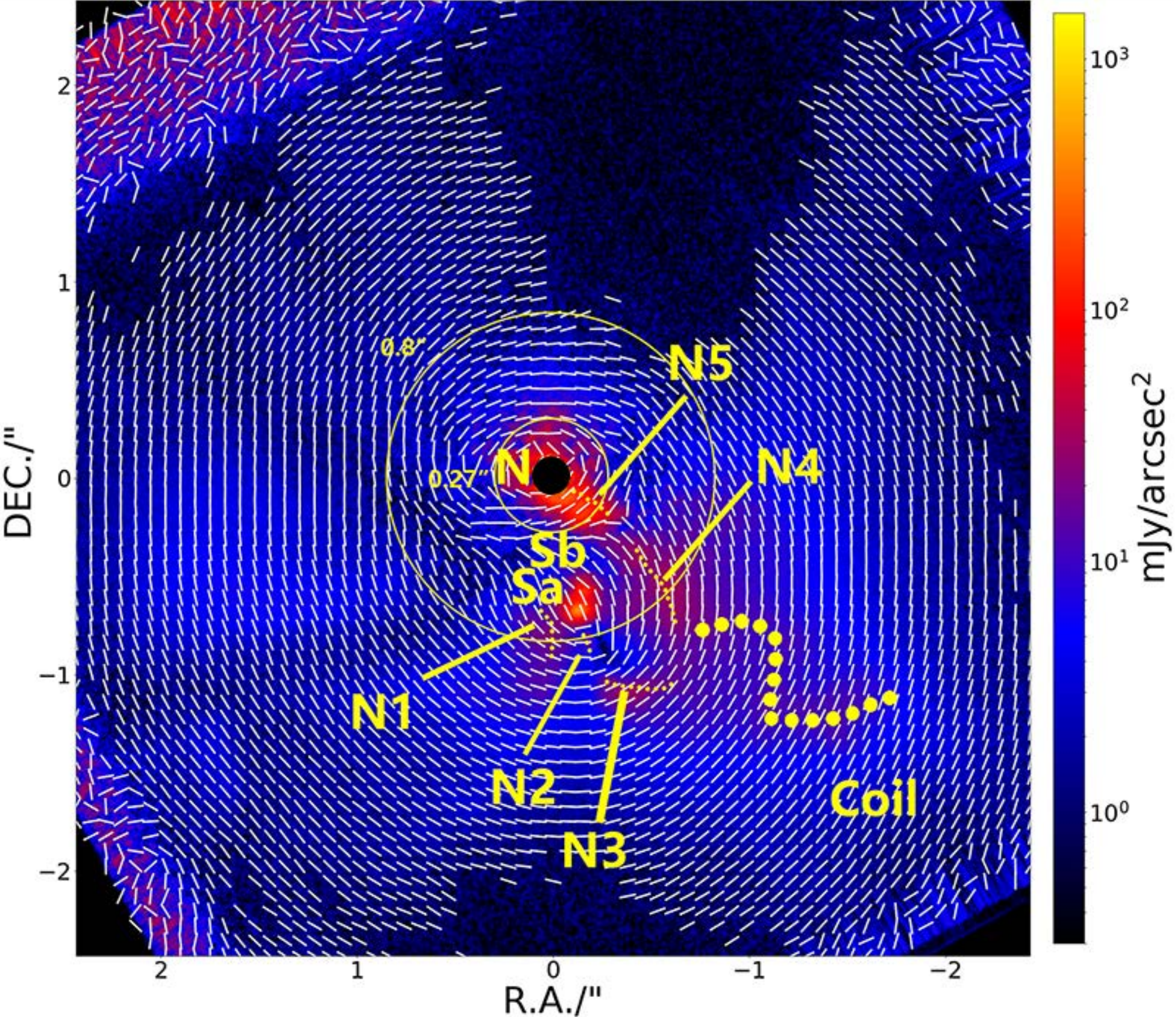}{0.5\textwidth}{a}
         \fig{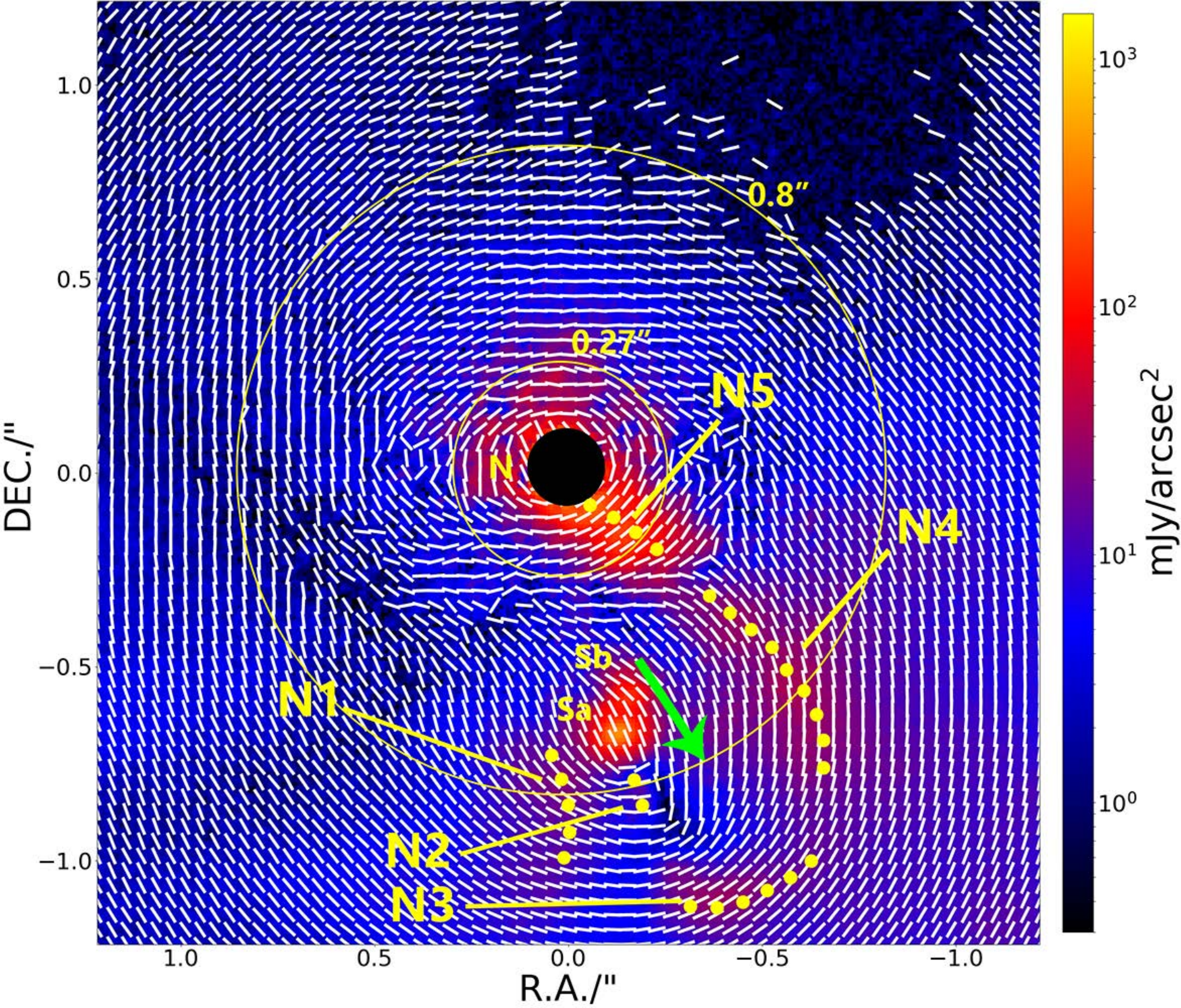}{0.5\textwidth}{b}
	}
	\caption{a: PI image of T Tau with polarization vector map overplotted. The vectors are calculated using the formula $\theta_p=0.5tan^{-1}(U/Q)$ in bins of 8 pixels. b: Central 2$\arcsec \times$2$\arcsec$ region of (a), calculated in bins of 4 pixels. Only the areas with signal-noise ratio larger than 4 are overplotted with vectors. The green arrow indicates the direction of aligned vectors.}
\end{figure}

\clearpage

\begin{figure}[ht!] 
    \figurenum{3}
    \begin{centering}
    \includegraphics[scale=0.6]{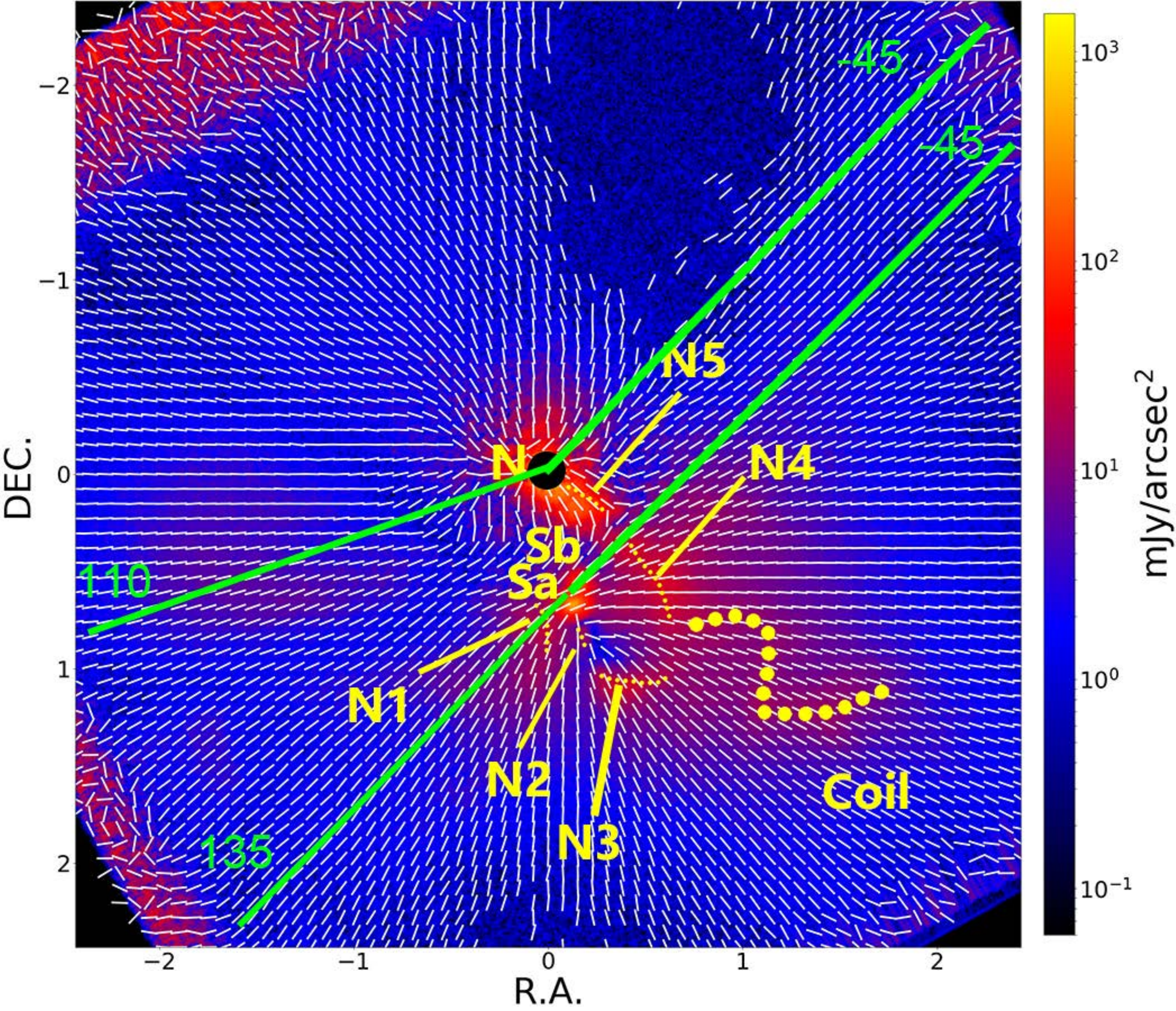}
    \par\end{centering}
    \caption{$PI$ image of T Tau with vectors perpendicular to the polarization vectors in Figure 2(a). From the directions of the vectors we can judge which area is illuminated by which star. The green lines and numbers indicate for T Tau N the vectors with position angles roughly from $-$45$^\circ$ to 110$^\circ$ pointing to T Tau N, and for T Tau S the vectors with position angles roughly from 135$^\circ$ to $-$45$^\circ$ (315$^\circ$).}
\end{figure}

\begin{figure}[ht!] 
    \figurenum{4}
    \begin{centering}
    \includegraphics[scale=0.4]{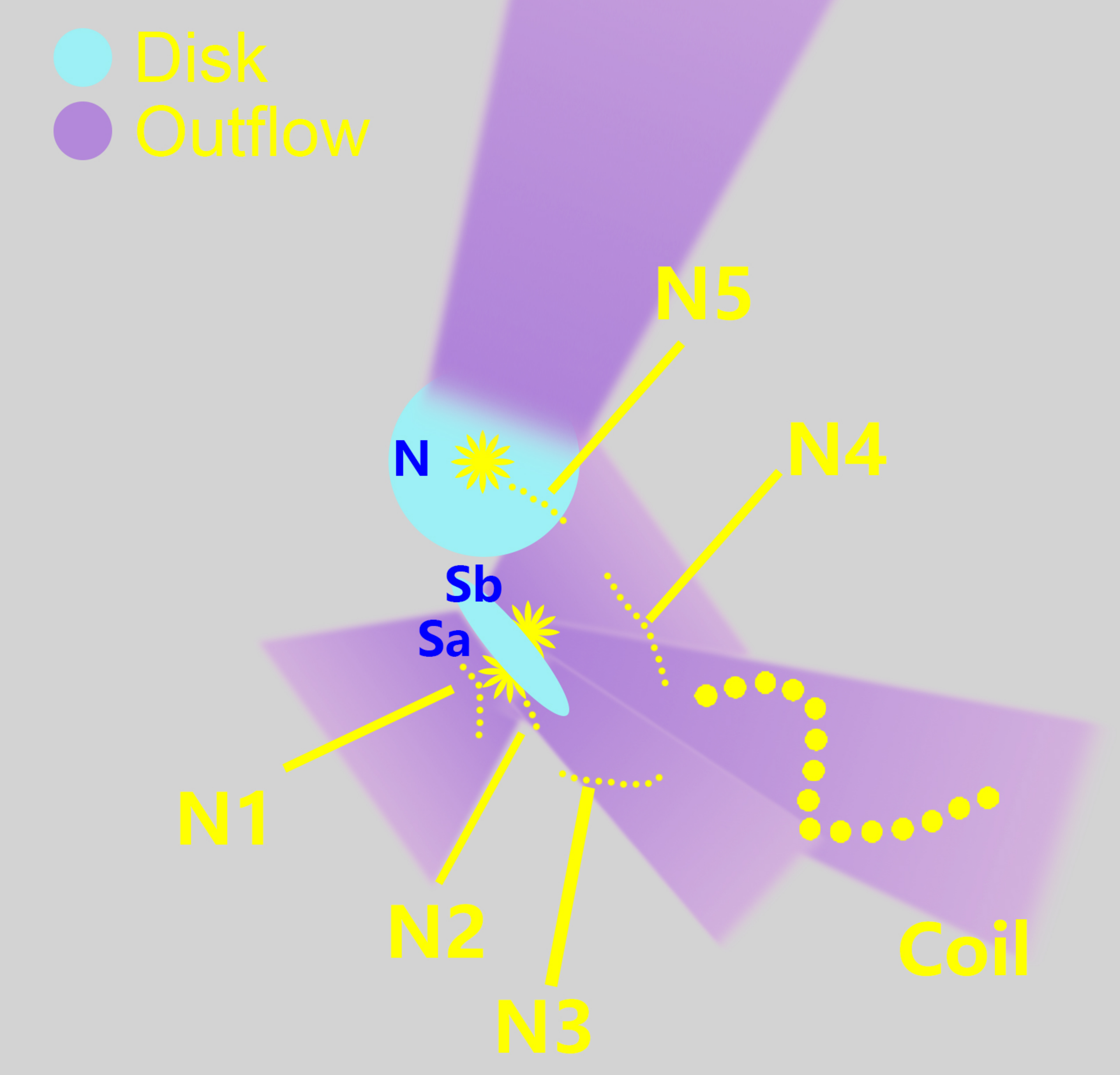}
    \par\end{centering}
    \caption{Illustration of suggested T Tau surrounding structures. The three stars are believed to be surrounded by envelopes (gray background). There are several possible outflows from this system (purple). In addition, both the T Tau N and T Tau S systems have disks (blue).}
\end{figure}

\listofchanges

\end{document}